\documentstyle[aps,prd,12pt,tighten,amssymb,amsmath,eqsecnum,cite]{revtex}

\def\bea{\begin{eqnarray}}
\def\eea{\end{eqnarray}}

\title{{\large{\bf Random Dirac Fermions: \\
The su$(N)$ Gauge Potential and ${\Bbb Z}_N$ Twists}}}
\vspace{0.5cm}
\author{Miraculous
J. Bhaseen\footnote{\vspace*{-0.4cm}bhaseen@thphys.ox.ac.uk} 
\\ \vspace{0.5cm}
{\small {\em Physics Department, University of Oxford, Theoretical Physics, \\ 1 Keble Road,
Oxford, OX1 3NP, U.K.}}}
\begin{document}
\maketitle

\begin{center}
\today
\end{center}
\vspace{0.5cm}

\begin{abstract}
We obtain the disorder averaged (critical) four-point correlation
functions for N species of  (two-dimensional
Euclidean) Dirac fermions subject to a (Gaussian) random
${\rm su}(N)$ gauge field. The replica approach and the strong disorder
approach yield identical results, as one might expect in the light of
recent developments in the $N=2$ case \cite{Bhaseen:marriage}. We establish a connection
with `dressed' ${\Bbb Z}_N$
twist fields in the $c=-2$ logarithmic conformal field theory (LCFT), thereby extending the
recent connection to `dressed' ${\Bbb Z}_2$ twists
\cite{Bhaseen:marriage,Ludwig:c=-2}. We draw attention to the fact
that `dressed' ${\Bbb
Z}_2$ twist operators have  previously appeared in the  description of half flux-quantum quasiholes
in the bulk excitations of the Haldane--Rezayi Quantum
Hall state \cite{Gurarie:Haldane}.
\end{abstract}




\section{Introduction}
A remarkable connection between the problem of
Dirac fermions in a random su$(2)$ gauge field and the $c=-2$
LCFT has recently emerged
\cite{Ludwig:c=-2,Bhaseen:marriage}. The disorder averaged correlation
functions of the so-called Q-matrix in the su$(2)$ random gauge
problem ---  from which one obtains correlation functions of the local density of
states (LDOS) \cite{Nersesyan:PRL72,Nersesyan:NPB438,Bhaseen:marriage} --- are those of ${\Bbb Z}_2$ twist fields in the $c=-2$ LCFT
`dressed' (or augmented) by the
exponential of a free bosonic field.  In this paper we demonstrate
that the connection holds beyond the $N=2$ case in that  disorder averaged
correlation functions in the su$(N)$ problem are obtained by
`dressing' those of
${\Bbb Z}_N$ twists.

The problem of random Dirac
fermions in an Abelian vector potential was originally introduced in
the context of the plateau transitions in the Integer Quantum Hall
Effect \cite{LudwigFSG}. The non-Abelian version of this problem
appeared in a treatment of disordered two-dimensional d-wave
superconductivity \cite{Nersesyan:PRL72,Nersesyan:NPB438} and has been
the subject of deeper investigation
\cite{JS:hidden,Mudry:twodcft,Komudtsvelik,Caux:exactmult,Caux2,Caux:disferm,Bhaseen:marriage,Bernard:perturbed}.
The $c=-2$ LCFT has also been studied extensively in  connection
with  dense polymers
\cite{Saleur:n=2,Ivashkevich:Polymers}, the Haldane--Rezayi Quantum
Hall state \cite{Gurarie:Haldane} and, somewhat more generally, as an instructive
example of a LCFT
\cite{Gurarie:Log,Kausch:Curiosities,Kausch:Symplectic,Gaberdiel:Local,Kogan:Boundarylog}.
That the random gauge field  problem and the $c=-2$ LCFT bear a
close relationship is intriguing, and warrants further investigation.

The structure of this paper is as follows: in section \ref{replicas}
we use the replica approach to study the four-point function of the
Q-matrix for the su$(N)$ problem. The
reliability of the replica approach is confirmed by
independent calculation in the limit of infinite disorder strength (Appendix
\ref{Strong}) as one might expect given the recent convergence of results for
three\footnote{Namely,  supersymmetry, replicas, and in the limit of
infinite disorder strength where a special treatment is possible.} distinct treatments of the $N=2$
case \cite{Bhaseen:marriage}. The  Q-matrix is governed by a
non-linear sigma model (familiar in the conventional theory of
localization) with the addition of a WZNW term (which ensures
criticality.) The Q-matrix correlation functions are obtained from the solution of the
su$(r)_N$ Knizhnik--Zamolodchikov equations (in the replica $r \rightarrow 0$ limit)
`dressed' by the exponential of a free bosonic field, 
as was first elucidated in \cite{Nersesyan:PRL72}. We demonstrate that this correlation function  assumes a simple
form in terms of a (N-species)
generalization of the (complete) elliptic integrals. The ordinary
(complete) elliptic integrals have played an instrumental r{\^o}le in
establishing the  connection between the random gauge field problem and
`dressed' ${\Bbb Z}_2$ twist operators of the $c=-2$ LCFT
\cite{Bhaseen:legendre,Ludwig:c=-2,Bhaseen:marriage}. In an analogous manner, the generalized
elliptic integrals enable one to establish a connection to `dressed'
${\Bbb Z}_N$ twist operators of the form appearing in
\cite{Dixon:orbifolds,Saleur:n=2}. Finally, we present
concluding remarks and a brief appendix on the most
significant identities which appear in the strong disorder treatment
of the su$(N)$ problem --- the fermionic `dressing' of the
$su(N)_{-2N}$ WZNW model \cite{Mudry:twodcft,Bernard:perturbed,Caux2}.
 
\section{Replicas and ${\Bbb Z}_N$ Twists}
\label{replicas}
As is discussed in section III of reference \cite{JS:hidden}, the four-point correlation
function of the Q-matrix (with all replica indices set to unity) reads 
\begin{gather}
\langle  Q_{11}(1)Q^\dagger_{11}(2)Q_{11}(3)Q^\dagger_{11}(4)\rangle
\sim \notag \\
|z_{13}z_{24}|^{-2/N^2}|z(1-z)|^{-2/N^2}\left[F_{11}+F_{12}+F_{21}+F_{22}\right]
\end{gather}
where $z=z_{12}z_{34}/z_{13}z_{24}$ and 
\begin{equation}
\label{singlevalued}
F_{ij}= F_{i}^{(1)}(z)F_{j}^{(2)}(\bar z) + F_{i}^{(2)}(z)F_{j}^{(1)}(\bar z)
\end{equation}
are single-valued combinations of the solutions to the  su$(r)_N$
Knizhnik--Zamolodchikov equations in the $r\rightarrow 0$ limit
\cite{Knizam:current,JS:hidden}. In the replica limit these equations
are of the form
\begin{equation}
Nz\frac{d F_1}{dz} = - F_2, \quad N(1 - z)\frac{d F_2}{dz} = F_1
\end{equation}
and admit  solutions in terms of the ordinary hypergeometric
functions\footnote{The prescribed normalization of these solutions ensures
the single-valuedness (monodromy invariance) of the combination (\ref{singlevalued}).}
\begin{subequations}
\begin{eqnarray}
F_{1}^{(1)} & = & \ _2F_{1}[-\tfrac{1}{N},\tfrac{1}{N};1;z] \\
F_{1}^{(2)} & = & (1-z)\ _2F_{1}[1-\tfrac{1}{N},1+\tfrac{1}{N};2;1-z] \\
F_{2}^{(1)} & = & \frac{z}{N}\
_2F_{1}[1-\tfrac{1}{N},1+\tfrac{1}{N};2;z] \\
F_{2}^{(2)} & = & N\ _2F_{1}[-\tfrac{1}{N},\tfrac{1}{N};1;1-z] 
\end{eqnarray}
\end{subequations}
It is a straightforward exercise in the application of the Gauss recursion
relations for the ordinary hypergeometric function to demonstrate that
\begin{equation}
F_{11}+F_{12}+F_{21}+F_{22} = \frac{4N}{\pi^2}\left[K_N(z)K_N(1-\bar z) + K_N(\bar z)K_N(1-z)\right]
\end{equation}
where we have defined the generalized  (complete) elliptic integral of the first
kind\footnote{The factor of $\pi/2$ ensures that $K_2$ agrees
with the usual definition of K. We note that the 
generalized (complete) elliptic integral of the second kind assumes
the form $E_N=\frac{\pi}{2}\ _2F_{1}[-\tfrac{1}{N},\tfrac{1}{N};1;z]$.}
\begin{equation}
K_N\equiv  \frac{\pi}{2}\ _2F_{1}[1-\tfrac{1}{N},\tfrac{1}{N};1;z].
\end{equation}
That is to say, upto overall normalization,
\begin{gather}
\langle Q_{11}(1)Q^\dagger_{11}(2)Q_{11}(3)Q^\dagger_{11}(4)\rangle
\sim \notag \\
|z_{13}z_{24}|^{-2/N^2}|z(1-z)|^{-2/N^2}\left[K_N(z)K_N(1-\bar z) +
K_N(1-z)K_N(\bar z)\right]. \label{Qcorr}
\end{gather}
We invite the reader to compare this four-point function with that of
${\Bbb Z}_N$ twist operators of the $c=-2$ LCFT as discussed  in the
context of dense polymers --- see equations (83) and (84) of reference
\cite{Saleur:n=2} and note the  small typing error in the prefactor:
\begin{gather}
\langle \sigma_{+}(1)\sigma_{-}(2)\sigma_{+}(3)\sigma_{-}(4)\rangle
\sim \notag \\
|z_{13}z_{24}|^{2k/N(1-k/N)}|z(1-z)|^{2k/N(1-k/N)}\left[
F_{k/N}(z)F_{k/N}(1-\bar z)+F_{k/N}(1-z)F_{k/N}(\bar z)\right] \label{sigma4pt}.
\end{gather}
Here $\sigma_{+}$ stands for the ${\Bbb Z}_{N}$ twist operator $\sigma_{k/N}$, $\sigma_{-}$ stands
for the ${\Bbb Z}_{N}$ anti-twist operator, both of conformal dimension
\begin{equation}
h=-\frac{1}{2}\frac{k}{N}\left(1-\frac{k}{N}\right)
\end{equation}
and where
\begin{equation}
F_{k/N}=\ _2F_{1}[1-\tfrac{k}{N},\tfrac{k}{N};1;z]. \label{FKN}
\end{equation}
For more details we refer the reader to \S 3.2 of \cite{Saleur:n=2}. In particular setting $k=1$ in equations (\ref{sigma4pt}) and
(\ref{FKN}) one obtains
\begin{gather}
\langle \sigma_{+}(1)\sigma_{-}(2)\sigma_{+}(3)\sigma_{-}(4)\rangle
\sim \notag \\
|z_{13}z_{24}|^{2/N(1-1/N)}|z(1-z)|^{2/N(1-1/N)}\left[
K_N(z)K_N(1-\bar z)+K_N(\bar z)K_N(1-z)\right] \label{sigmas}
\end{gather}
It is readily seen that equations (\ref{Qcorr}) and (\ref{sigmas}) coincide upto
a simple prefactor to which we now turn our attention. Adopting the
conventions appearing in chapter 9 of \cite{Francesco:CFT} we consider the
four-point correlation functions of the exponentials of free bosonic fields
\begin{equation}
\nu_\alpha=e^{i\sqrt 2 \alpha \varphi}
\end{equation}
governed by the action
\begin{equation}
S=\frac{1}{8\pi} \int d^2x \,\partial_\mu\varphi \partial^\mu\varphi.
\end{equation}
and of conformal dimension $h=\alpha^2$. The four-point function of
such fields is 
\begin{equation}
\langle
\nu_{\alpha}(1)\nu_{-\alpha}(2)\nu_{\alpha}(3)\nu_{-\alpha}(4)\rangle =
|(z_{13}z_{24})|^{-4\alpha^2}|z(1-z)|^{-4\alpha^2}
\end{equation}
The necessary compensating prefactor between equations (\ref{Qcorr}) and (\ref{sigmas}) is obtained by setting $\alpha^2=1/2N$.
That is to say, one obtains the following identification between the
Q-matrix and the `dressed' ${\Bbb Z}_{N}$ twist operator of $c=-2$:
\begin{eqnarray}
Q \sim  e^{i\varphi/\sqrt N}\,\sigma_{+}, \quad Q^\dagger  \sim  e^{-i\varphi/\sqrt N}\,\sigma_{-}
\end{eqnarray}
Indeed, the conformal dimensions of the bosonic exponent and the
${\Bbb Z}_N$ twist
operator (with k=1) add up to
\begin{equation}
\frac{1}{2N}-\frac{1}{2}\frac{1}{N}\left(1-\frac{1}{N}\right)=\frac{1}{2N^2}
\end{equation}
which is the known conformal dimension of the
Q-matrix/LDOS\cite{Nersesyan:PRL72}.
\section{Conclusions}
In this paper we have discussed the explicit connection between Dirac fermions in a random su$(N)$ gauge field and  ${\Bbb Z}_N$ twists
 in the $c=-2$ LCFT. This generalizes  the `dressing' of the $h=-1/8$
 twist field of relevance in the $N=2$ case
 \cite{Bhaseen:marriage,Ludwig:c=-2}. We note that we have not discussed the
 $N>2$ analogue of `dressing' the $h=3/8$ twist field,  but one
 anticipates a connection to the excited twist fields $\tau_{k/N}$
 disussed in \cite{Saleur:n=2}.
 
Given that dressed ${\Bbb Z}_2$ twist operators are deemed respsonsible for the creation
of half flux-quantum quasiholes in the bulk excitations of the Haldane--Rezayi
Quantum Hall state --- see section
V of reference \cite{Gurarie:Haldane} --- one might expect a ${\Bbb
Z}_N$ generalization to shed some light on the su$(N)$ random gauge
field problem.
	
These issues are under current investigation.
\section*{Acknowledgements}The author would like to thank J.-S. Caux, I. I. Kogan, A. M. Tsvelik, for stimulating
and valuable discussions. The author would also like to thank EPSRC for
financial support.
\appendix
\section{Strong Disorder}
\label{Strong}
In the limit of strong disorder one may obtain the correlation
functions of the Q-matrix by a fermionic `dressing' of the solutions
to the su$(N)_{-2N}$ Knizhnik--Zamolodchikov equations
\cite{Mudry:twodcft,Bernard:perturbed,Caux2}. These solutions may be
written in the form  $F_{i}^{(a)}=z^{1-1/N^2}(1-z)^{1-1/N^2}{\mathcal
F}_{i}^{(a)}$ where 
\begin{subequations}
\begin{eqnarray}
{\mathcal F}_{1}^{(1)} & = & (1-z) \
_2F_1[2-\tfrac{1}{N},2+\tfrac{1}{N};2;z]\\
{\mathcal F}_{1}^{(2)} & = &  (1-z)   \
_2F_1[2-\tfrac{1}{N},2+\tfrac{1}{N};3;1-z]\\
{\mathcal F}_{2}^{(1)} & = & -\frac{1}{2N}z \
_2F_1[2-\tfrac{1}{N},2+\tfrac{1}{N};3;z]\\
{\mathcal F}_{2}^{(2)} & = & -2N z  \
_2F_1[2-\tfrac{1}{N},2+\tfrac{1}{N};2;1-z].
\end{eqnarray}
\end{subequations}
Dressing with fermions leads
one to consider single-valued combinations of the functions 
\begin{equation}
g^{(a)}=N{\mathcal F}_{1}^{(a)}+{\mathcal
F}_2^{(a)}+\frac{z}{1-z}[N{\mathcal F}_2^{(a)}+{\mathcal F}_1^{(a)}].
\end{equation}
A straightforward but tedious application of the Gauss recursion relations for
hypergeometric functions allows one to recover the generalized
(complete) elliptic integrals:
\begin{subequations}
\begin{eqnarray}
g^{(1)}  =    \frac{2N}{\pi(1-z)}K_N(z), \quad
g^{(2)}  =   -\frac{4N^2}{\pi(1-z)}K_N(1-z)
\end{eqnarray}
\end{subequations}
In this manner one is able to recover the results of the replica approach.

\end{document}